\documentclass[a4paper]{PoS}

\title{Three Dirac operators on two architectures with one piece of code and no hassle}

\ShortTitle{Three Dirac operators on two architectures}

\author{\speaker{Stephan D\"urr}\\ 
        University of Wuppertal, D-42119 Wuppertal, Germany\\
        IAS/JSC Forschungszentrum J\"ulich, D-52425 J\"ulich, Germany\\
        E-mail: \email{durr\,(AT)\,itp$\mbox{.}$unibe$\mbox{.}$ch}}

\abstract{A simple minded approach to implement three discretizations of the Dirac operator (staggered, Wilson, Brillouin) on two architectures (KNL and core\,i7) is presented.
The idea is to use a high-level compiler along with OpenMP parallelization and SIMD pragmas, but to stay away from cache-line optimization and/or assembly-tuning.
The implementation is for $N_v$ right-hand-sides, and this extra index is used to fill the SIMD pipeline.
On one KNL node single precision performance figures for $N_c=3$, $N_v=12$ read 475 Gflop/s, 345 Gflop/s, and 790 Gflop/s for the three discretization schemes, respectively.}

\FullConference{The 36th Annual International Symposium on Lattice Field Theory - LATTICE2018\\
        22-28 July, 2018\\
        Michigan State University, East Lansing, Michigan, USA.}

\renewcommand{\dag}{^\dagger}

\newcommand{\nab}{\nabla}
\newcommand{\lap}{\triangle}

\newcommand{\ga}{\gamma}
\newcommand{\de}{\delta}

\newcommand{\et}{\eta}

\newcommand{\la}{\lambda}
\newcommand{\rh}{\rho}

\newcommand{\si}{\sigma}

\newcommand{\bdm}{\begin{displaymath}}
\newcommand{\edm}{\end{displaymath}}
\newcommand{\bea}{\begin{eqnarray}}
\newcommand{\eea}{\end{eqnarray}}
\newcommand{\beq}{\begin{equation}}
\newcommand{\eeq}{\end{equation}}

\newcommand{\mr}{\mathrm}

\begin{document}


\section{Introduction}

Recent years brought plenty of machines with peak performances in the multi-petaflop/s range, but it gets increasingly difficult to achieve good strong-scaling behavior with actual scientific codes.
Lattice QCD is still in a fortunate position to harness these capabilities~\cite{Boyle:2017wul,Rago:2017pyb,Lin:2018}, as it does not require any run-time dependent data structures.
On the other hand, an un-optimized non-parallel code tends to have $O(10^5)$ lines.
This means that significant human resources must be spent to parallelize a lattice code and to obtain good performance figures on a given architecture.

Ideally one would have a piece of code, written in a high-level language, which parallelizes and reaches decent (read: non-optimal but non-disastrous) performance upon compilation on a given new architecture.
In these proceedings I report on an attempt to do this on the single-node level, based on OpenMP (OMP) threads and again OpenMP pragmas for SIMD-pipelining.
I concentrate on the part which takes most time in actual computations -- the matrix-times-vector operation for a given Dirac operator, considering the Susskind (``staggered''), Wilson and Brillouin varieties.


\section{Vector layout options}

The routines are written in Fortran\,2008, using the stride notation, as this allows for compact source files (like in matlab).
The gauge field $U$ is defined as a 7-dimensional array through {\tt complex(kind=sp),dimension(Nc,Nc,4,Nx,Ny,Nz,Nt)\,::\,U},
with parameters like {\tt Nc=3} and {\tt sp,dp} (for single and double precision, respectively) specified at compile time.
Hence {\tt U(:,:,3,x,y,z,t)} defines $N_c^2$ complex numbers, arranged contiguously in memory.

With Wilson-type vectors arranged in blocks of $N_v$ right-hand sides, the layout options include
{\tt vec(Nc,4,Nv,...)}, {\tt vec(4,Nc,Nv,...)}, {\tt vec(Nc,Nv,4,..)}, {\tt vec(Nv,Nc,4,..)},
{\tt vec(4,Nv,Nc,...)}, {\tt vec(Nv,4,Nc,...)}, where the dots stand for {\tt Nx*Ny*Nz*Nt}.
This restriction of having the space-time index as the slowest (right-most) index precludes sophisticated SIMD strategies (see Refs.\,\cite{Boyle:2017wul,Rago:2017pyb}), but it may facilitate the use of PGAS concepts (see below).
With Susskind-type vectors arranged in blocks of $N_v$ right-hand sides, the layout options under the same restriction are
{\tt suv(Nc,Nv,Nx*Ny*Nz*Nt)}, {\tt suv(Nv,Nc,Nx*Ny*Nz*Nt)}.

Our task is to optimize the performance under the self-imposed set of restrictions.
An important ingredient in the code is that all contributions to the ``out'' vector are collected in the thread-private variable {\tt site}, which for each space-time index is written \emph{once}.
This avoids write collisions among threads in a natural way.
We have the same 6 layout options as for {\tt vec} to define {\tt site} as an array of dimension 3 in the Wilson case (or the same 2 options in the staggered case).


\section{Staggered kernel details and performance}

The Susskind (``staggered'') Dirac operator is defined through
\beq
D_\mr{S}(x,y)=\sum_{\mu} \et_\mu(x)\,\frac{1}{2}\,[V_{\mu}(x)\de_{x+\hat\mu,y}-V_{\mu}\dag(x-\hat\mu)\de_{x-\hat\mu,y}]
\eeq
with $\et_1(x)=1$, $\et_2(x)=(-1)^{x_1}$, $\et_3(x)=(-1)^{x_1+x_2}$, $\et_4(x)=(-1)^{x_1+x_2+x_3}$.
Here $V_\mu(x)$ represents a smeared version of the (original) gauge link $U_\mu(x)$, i.e.\ a gauge-covariant parallel transporter from $x+\hat\mu$ to $x$, to reduce taste-symmetry breaking.
From a HPC viewpoint, a clear advantage of this operator with precomputed $V_\mu$ is that its stencil is restricted to sites which are at most one hop away.
Still, it is not trivial to reach an acceptable performance on a many-core architecture~\cite{DeTar:2016ndn,DeTar:2018pyj}.

\begin{figure}[tb!]
\includegraphics[width=0.5\textwidth]{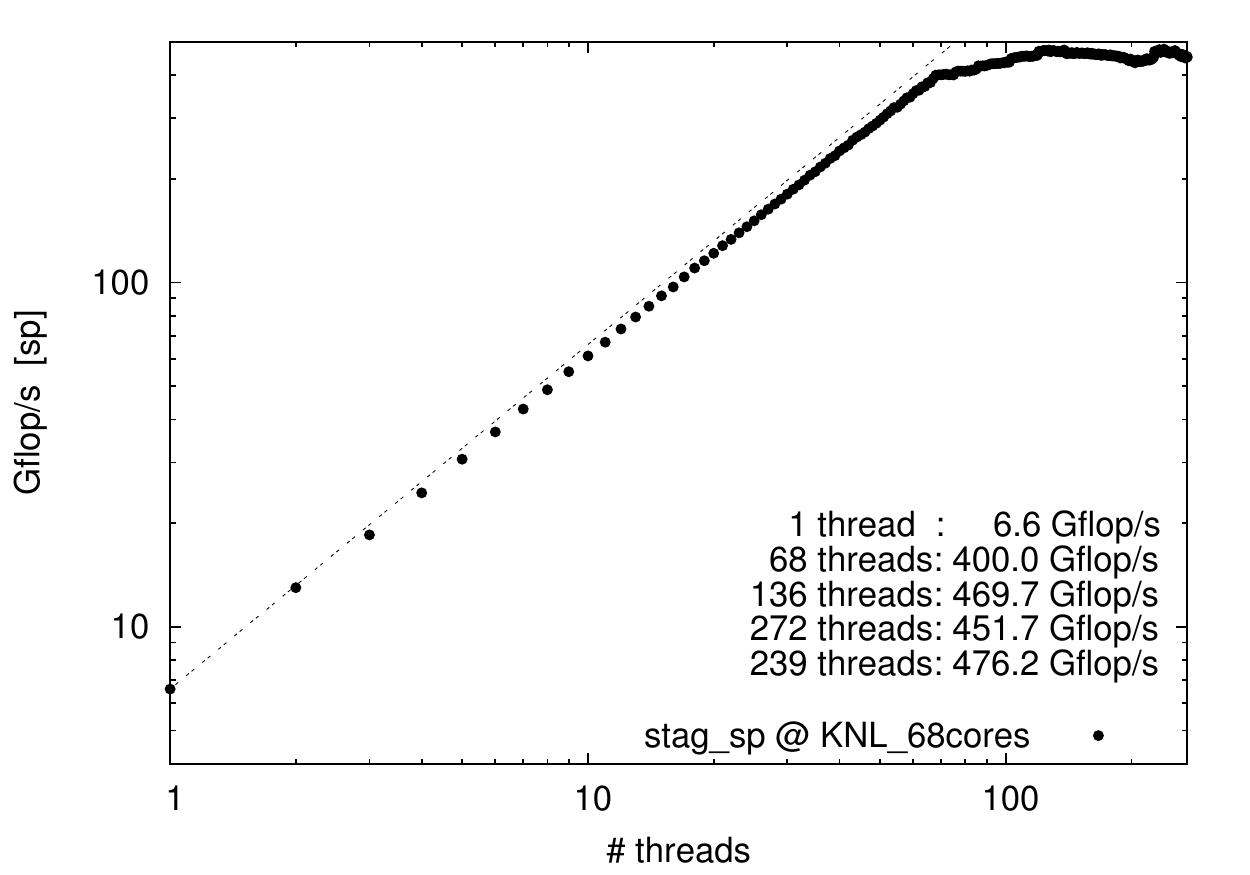}%
\includegraphics[width=0.5\textwidth]{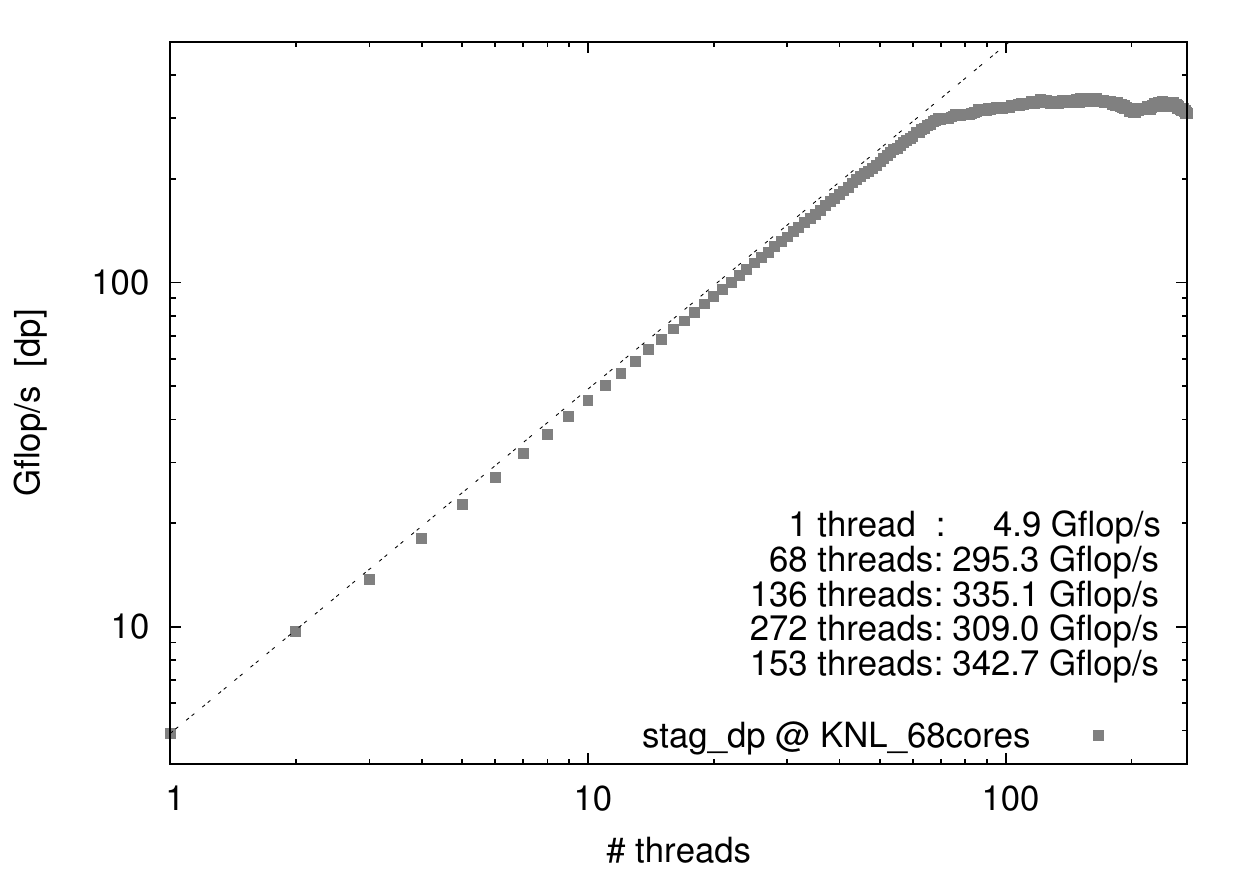}%
\\[2mm]
\includegraphics[width=0.5\textwidth]{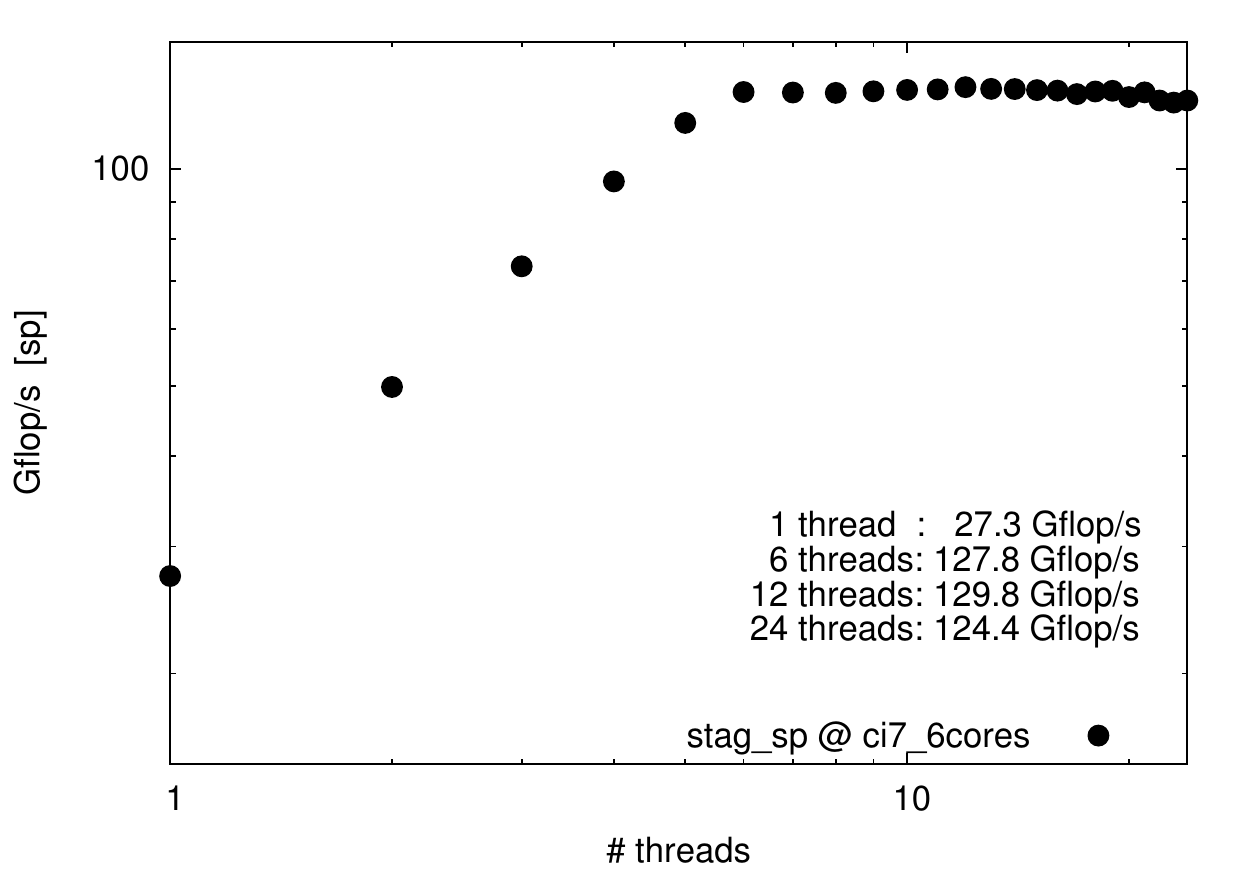}%
\includegraphics[width=0.5\textwidth]{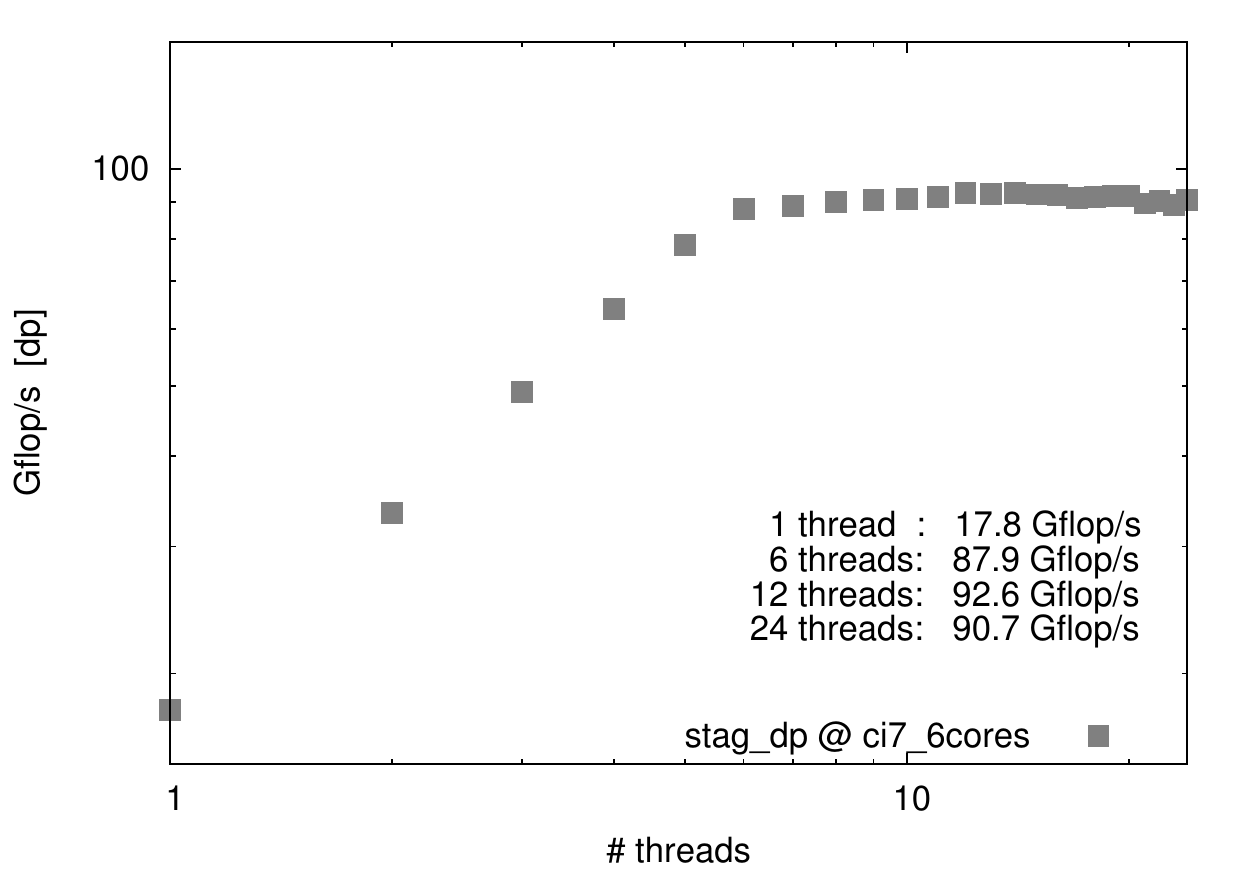}%
\caption{\label{fig:stag}
Log-log scaling plot of the single processor performance in Gflop/s versus the number of OMP threads for the staggered Dirac operator in {\tt sp} (left) and {\tt dp} (right).
The top panels feature a KNL processor with 68 cores, the bottom panels a Broadwell chip with 6 cores. The lattice size is $34^3\times68$.}
\end{figure}

In our framework we have 2 options for the {\tt suv} layout, 2 options for {\tt site}, and 2 reasonable loop nestings.
It is thus possible to implement all these options, and to compare the timings.
For one choice (the one performing best on the KNL architecture) thread scaling results are shown in Fig.\,\ref{fig:stag}.
Performance on the Broadwell architecture seems far less sensitive to these choices.


\section{Wilson kernel details and performance}

The Wilson Dirac operator is defined through
\beq
D_\mr{W}(x,y)=\sum_\mu \ga_\mu \nab_\mu^\mr{std}(x,y)
-\frac{a}{2}\lap^\mr{std}(x,y)+m_0\de_{x,y}
-\frac{c_\mr{SW}}{2}\sum_{\mu<\nu}\si_{\mu\nu}F_{\mu\nu}\de_{x,y}
\;,
\label{def_wils}
\eeq
where $\nab_\mu^\mr{std}$ is a 2-point discretization of the covariant derivative
\beq
a\nab_\mu^\mr{std}(x,y)=\frac{1}{2}\,[V_{\mu}(x)\de_{x+\hat\mu,y}-V_{-\mu}(x)\de_{x-\hat\mu,y}]
\eeq
and $\lap^\mr{std}$ is a 9-point discretization of the covariant Laplacian (sum over 4 pos.\ and 4 neg.\ indices)
\beq
a^2\lap^\mr{std}(x,y)=-\,8\,\de_{x,y}+1\sum\nolimits_{\mu}V_\mu(x)\de_{x+\hat\mu,y}
\;.
\eeq
Also this operator is HPC friendly, since its stencil contains at most 1-hop terms.

\begin{figure}[tb!]
\includegraphics[width=0.5\textwidth]{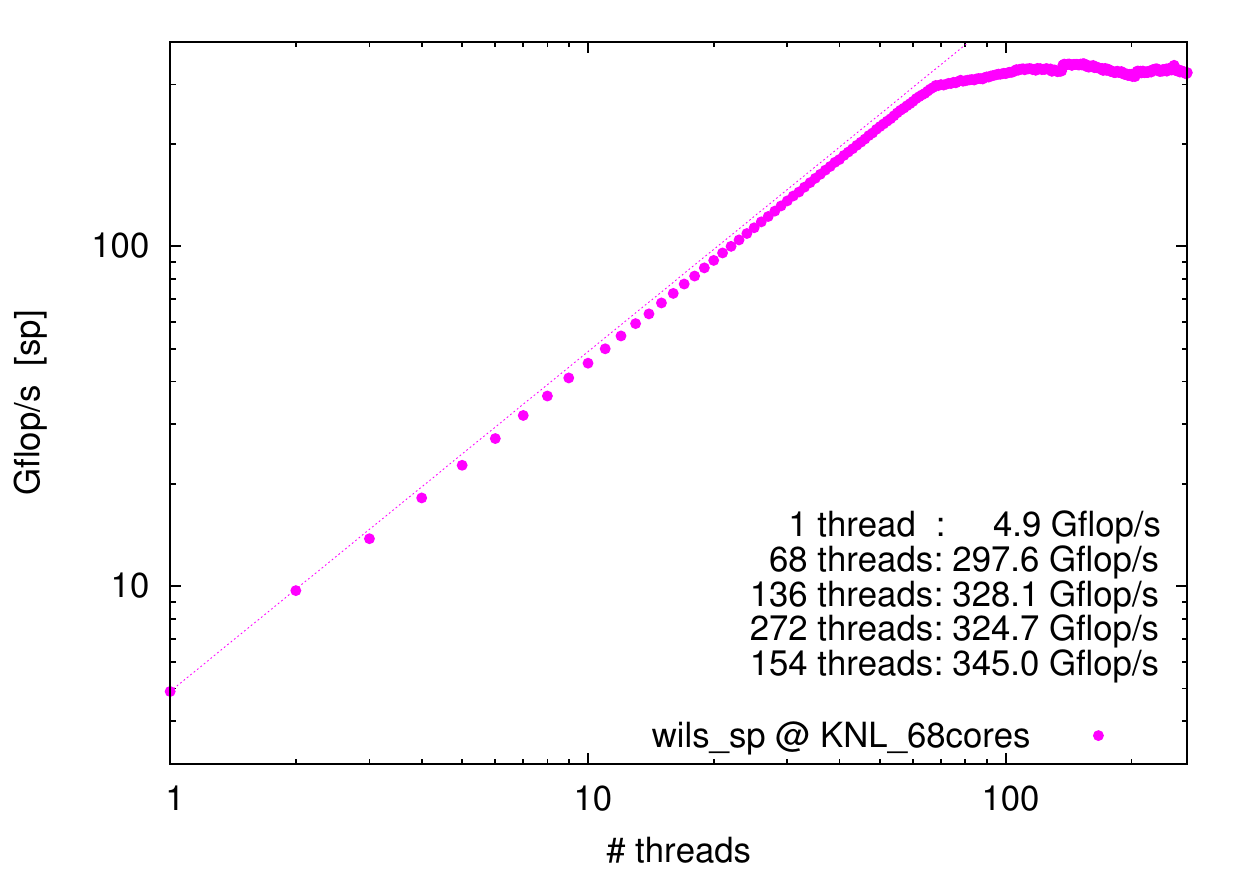}%
\includegraphics[width=0.5\textwidth]{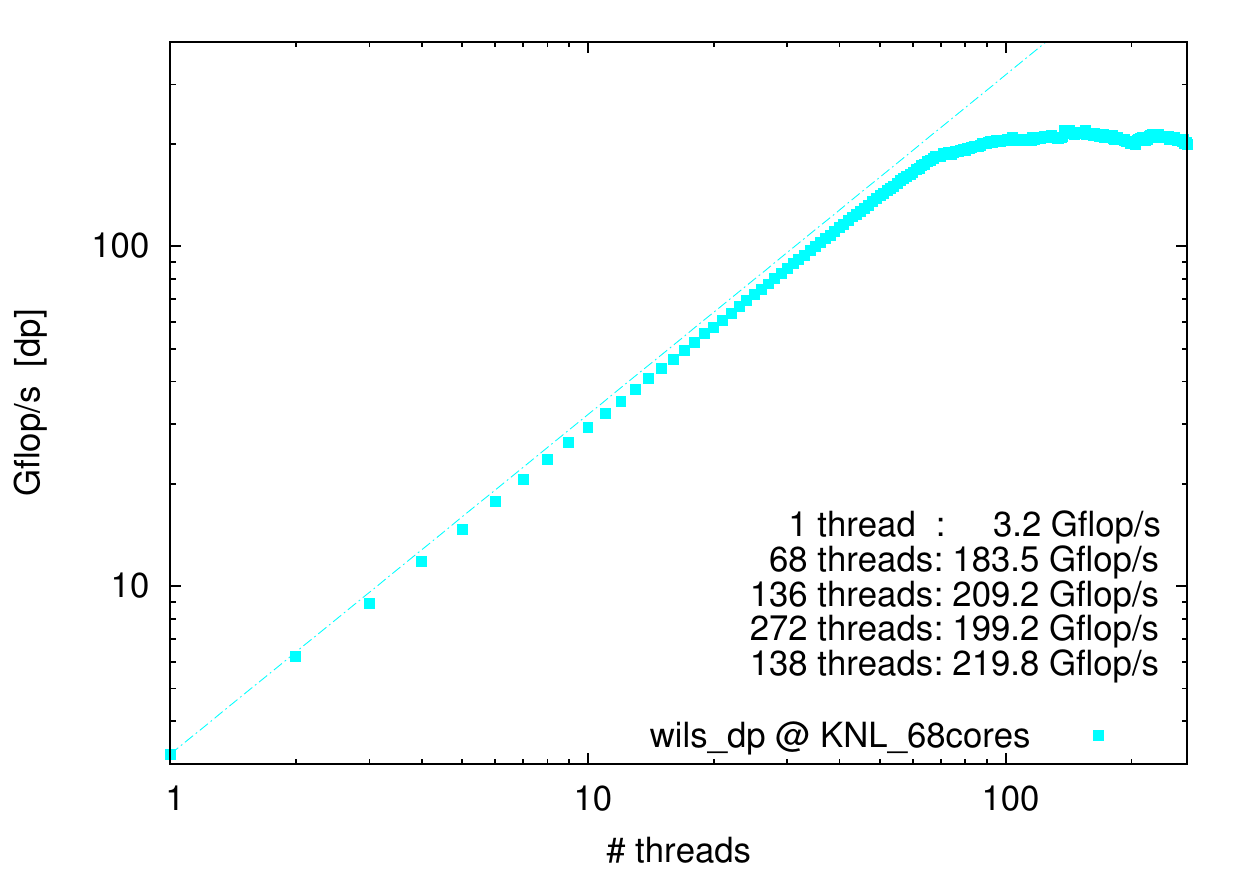}%
\\[2mm]
\includegraphics[width=0.5\textwidth]{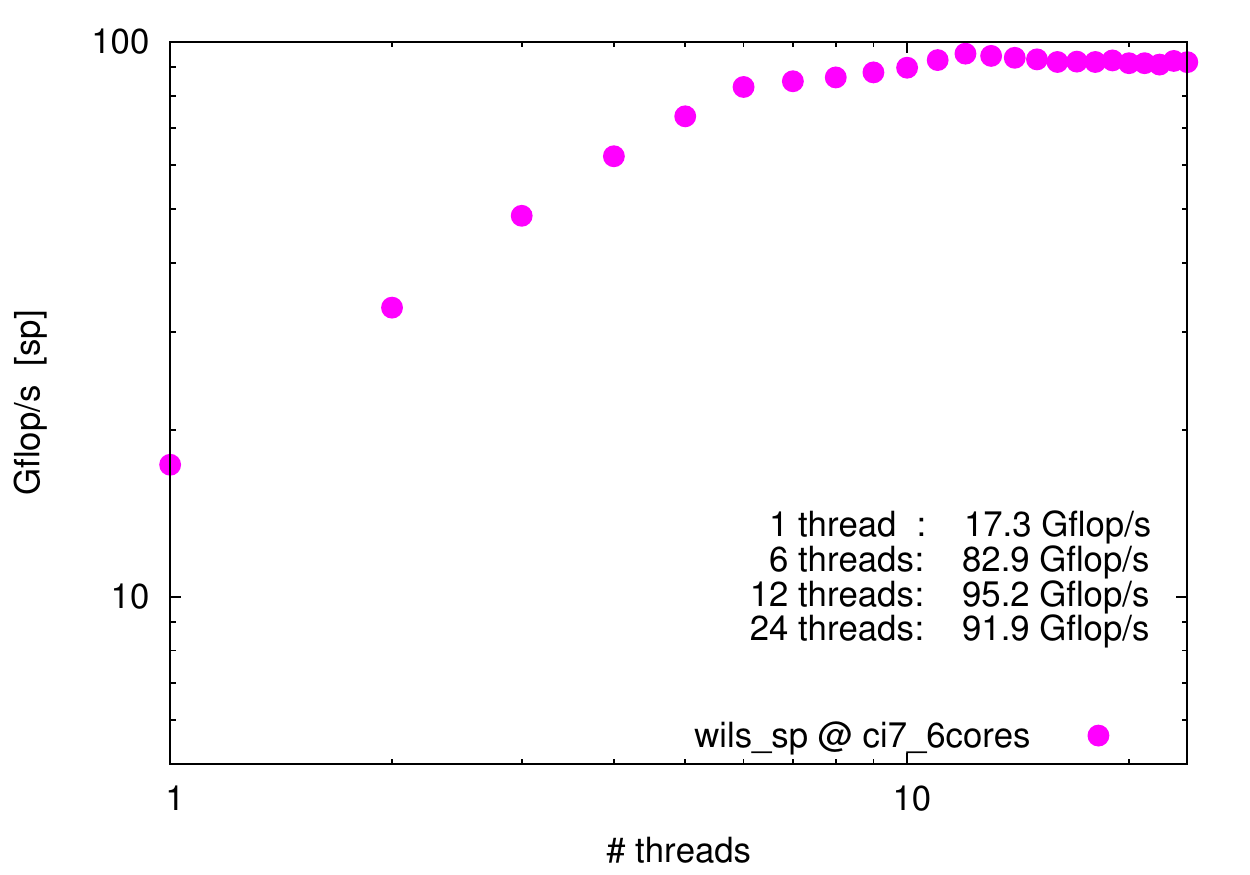}%
\includegraphics[width=0.5\textwidth]{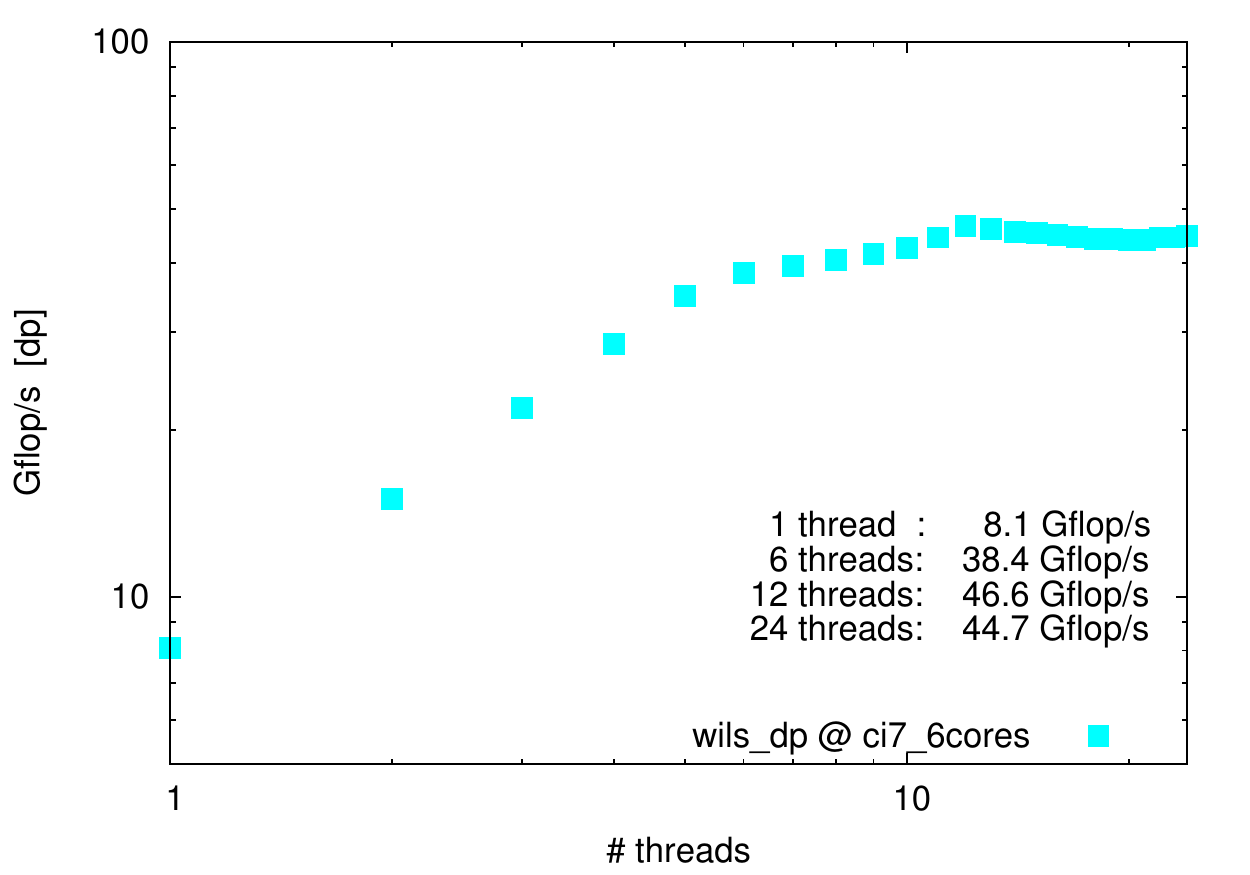}%
\caption{\label{fig:wils}
Log-log scaling plot of the single processor performance in Gflop/s versus the number of OMP threads for the Wilson Dirac operator at $c_\mr{SW}=0$ in {\tt sp} (left) and {\tt dp} (right).
The top panels feature a KNL processor with 68 cores, the bottom panels a Broadwell chip with 6 cores. The lattice size is $34^3\times68$.}
\end{figure}

In our framework we have 6 options for the {\tt vec} layout, times 6 options for {\tt site}, times a few reasonable loop nestings.
For the standard Laplacian $\lap^\mr{std}$ in the Wilson operator the latter set of options is 4, resulting in a total of 144 routines.
Evidently, this brings some limitations to the ``best of breed'' ansatz, since it may be a little awkward to test each routine with each possible thread count (e.g.\ 1 through 272 on the KNL).
In practice, it seems reasonable to restrict the selection process to just a few thread counts (e.g.\ 68, 136, and 272 on the KNL).

Full thread scaling results for one such choice are displayed in Fig.\,\ref{fig:wils}.
Similar to Fig.\,\ref{fig:stag}, we see an almost linear increase in performance up to 68 threads on the KNL.
This is followed by another linear gain (albeit with a smaller slope) up to 136 threads.
Beyond that point results wiggle a bit out to 272 threads.
The maximum appears in a number of threads (154 in {\tt sp}, 138 in {\tt dp}) which seems hard to predict.
Again, the code seems to perform (without any change) reasonably well on the Broadwell architecture, too.
Having more than 2 threads per core does not enhance performance, but the good news is that it's not really detrimental either.
Next steps of performance tuning would naturally include eo-decomposition and gauge compression (from $N_c$ to $N_c-1$ columns).


\section{Brillouin kernel details and performance}

The Brillouin Dirac operator is defined through~\cite{Durr:2010ch}
\beq
D_\mr{B}(x,y)=\sum_\mu \ga_\mu \nab_\mu^\mr{iso}(x,y)
-\frac{a}{2}\lap^\mr{bri}(x,y)+m_0\de_{x,y}
-\frac{c_\mr{SW}}{2}\sum_{\mu<\nu}\si_{\mu\nu}F_{\mu\nu}\de_{x,y}
\;,
\label{def_bril}
\eeq
where the isotropic derivative $\nab_\mu^\mr{iso}$ is a 54-point discretization of the covariant derivative
\bea
a\nab_\mu^\mr{iso}(x,y)&=&\rh_1\,[W_{\mu}(x)\de_{x+\hat\mu,y}-W_{-\mu}(x)\de_{x-\hat\mu,y}]
\nonumber\\
&+&\rh_2\sum\nolimits_{\neq(\nu;\mu)}[W_{\mu+\nu}(x)\de_{x+\hat\mu+\hat\nu,y}-(\mu\to-\mu)]
\nonumber\\
&+&\rh_3\sum\nolimits_{\neq(\nu,\rh;\mu)}[W_{\mu+\nu+\rh}(x)\de_{x+\hat\mu+\hat\nu+\hat\rh,y}-(\mu\to-\mu)]
\nonumber\\
&+&\rh_4\sum\nolimits_{\neq(\nu,\rh,\si;\mu)}[W_{\mu+\nu+\rh}(x)\de_{x+\hat\mu+\hat\nu+\hat\rh+\hat\si,y}-(\mu\to-\mu)]
\label{def_der}
\eea
and the Brillouin Laplacian $\lap^\mr{bri}$ is a 81-point discretization of the covariant Laplacian
\bea
a^2\lap^\mr{bri}(x,y)=\la_0\,\de_{x,y}
&+&\la_1\sum\nolimits_{\mu}W_{\mu}(x)\de_{x+\hat\mu,y}
\nonumber\\
&+&\la_2\sum\nolimits_{\neq(\mu,\nu)}W_{\mu+\nu}(x)\de_{x+\hat\mu+\hat\nu,y}
\nonumber\\
&+&\la_3\sum\nolimits_{\neq(\mu,\nu,\rh)}W_{\mu+\nu+\rh}(x)\de_{x+\hat\mu+\hat\nu+\hat\rh,y}
\nonumber\\
&+&\la_4\sum\nolimits_{\neq(\mu,\nu,\rh,\si)}W_{\mu+\nu+\rh+\si}(x)\de_{x+\hat\mu+\hat\nu+\hat\rh+\hat\si,y}
\label{def_lap}
\eea
with $(\rh_1,\rh_2,\rh_3,\rh_4)\equiv(64,16,4,1)/432$ and $(\la_0,\la_1,\la_2,\la_3,\la_4)\equiv(-240,8,4,2,1)/64$ respectively.
In (\ref{def_der}) the last sum extends over (pos.\ and neg.) indices $(\nu,\rh,\si)$ which are mutually unequal and different from $\mu$.
In (\ref{def_lap}) the last sum extends over indices $(\mu,\nu,\rh,\si)$ which are pairwise unequal.
In these formulas $W_\mr{dir}(x)$ denotes a link in direction ``dir'' which may be on-axis (dir=$\mu$) or off-axis with length $\sqrt{2}$ (dir=$\mu\nu$) or $\sqrt{3}$ (dir=$\mu\nu\rh$) or $\sqrt{4}$ (dir=$\mu\nu\rh\si$).
More details are given in Tab.\,\ref{tab:off-axis}.

\begin{table}[!tb]
\begin{tabular}{|@{\,}c@{\,\,}c@{\,\,}l@{\quad}l@{\,}|}
\hline
\#hop & \#terms & \#paths & formula \\
\hline
1 & 8 & 1!=1  & $W_{\mu}(x)=V_{\mu}(x)$ (smeared link, $\mu\!\in\!\{\pm1,\pm2,\pm3,\pm4\}$)\\
2 &24 & 2!=2  & $W_{\mu+\nu}(x)=\frac{1}{2}[V_\mu(x)V_\nu(x\!+\!\hat\mu)+\mr{perm}]$\\
3 &32 & 3!=6  & $W_{\mu+\nu+\rh}(x)=\frac{1}{6}[V_\mu(x)V_\nu(x\!+\!\hat\mu)V_\rh(x\!+\!\hat\mu\!+\!\hat\nu)+\mr{perms}]$\\
4 &16 & 4!=24 & $W_{\mu+\nu+\rh+\si}(x)=\frac{1}{24}[V_\mu(x)V_\nu(x\!+\!\hat\mu)V_\rh(x\!+\!\hat\mu\!+\!\hat\nu)V_\si(x\!+\!\hat\mu\!+\!\hat\nu\!+\!\hat\si)+\mr{perms}]$\\
\hline
\end{tabular}
\caption{\label{tab:off-axis}
Overview of the set of off-axis links $W_\mr{dir}(x)$, with lengths ranging from 1 to 4 hops.
Given a site $x$, 81 directions are possible, but one is trivial, and the remaining 80 can be reduced to 40 based on $W_\mr{-dir}^{}(x)=W_\mr{dir}\dag(x-\mr{dir})$.
In the code $W$ is precomputed and stored in the array {\tt W(Nc,Nc,40,Nx,Ny,Nz,Nt)}.
Note that for 36 of the 40 directions the entry is not special unitary, and no gauge compression is possible.}
\end{table}

\begin{figure}[tb!]
\includegraphics[width=0.5\textwidth]{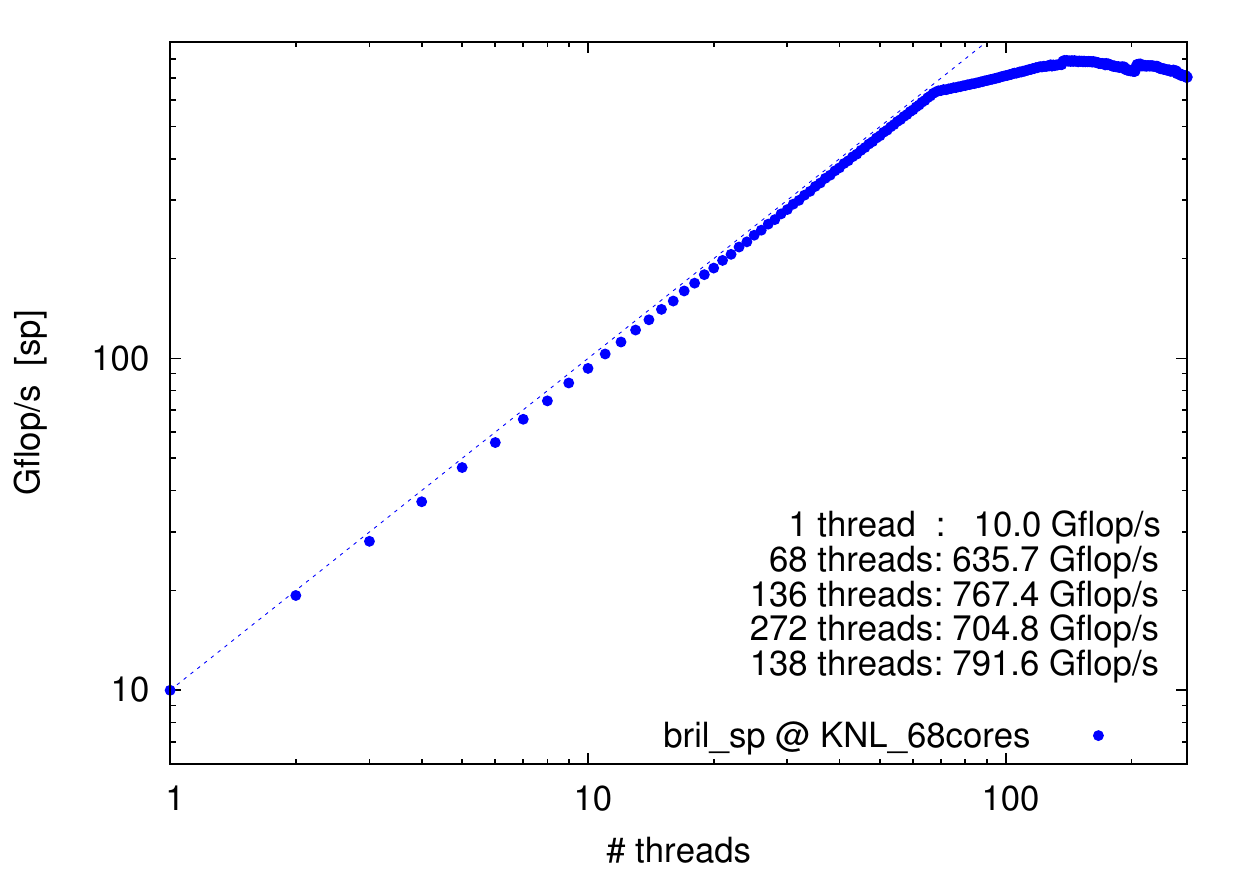}%
\includegraphics[width=0.5\textwidth]{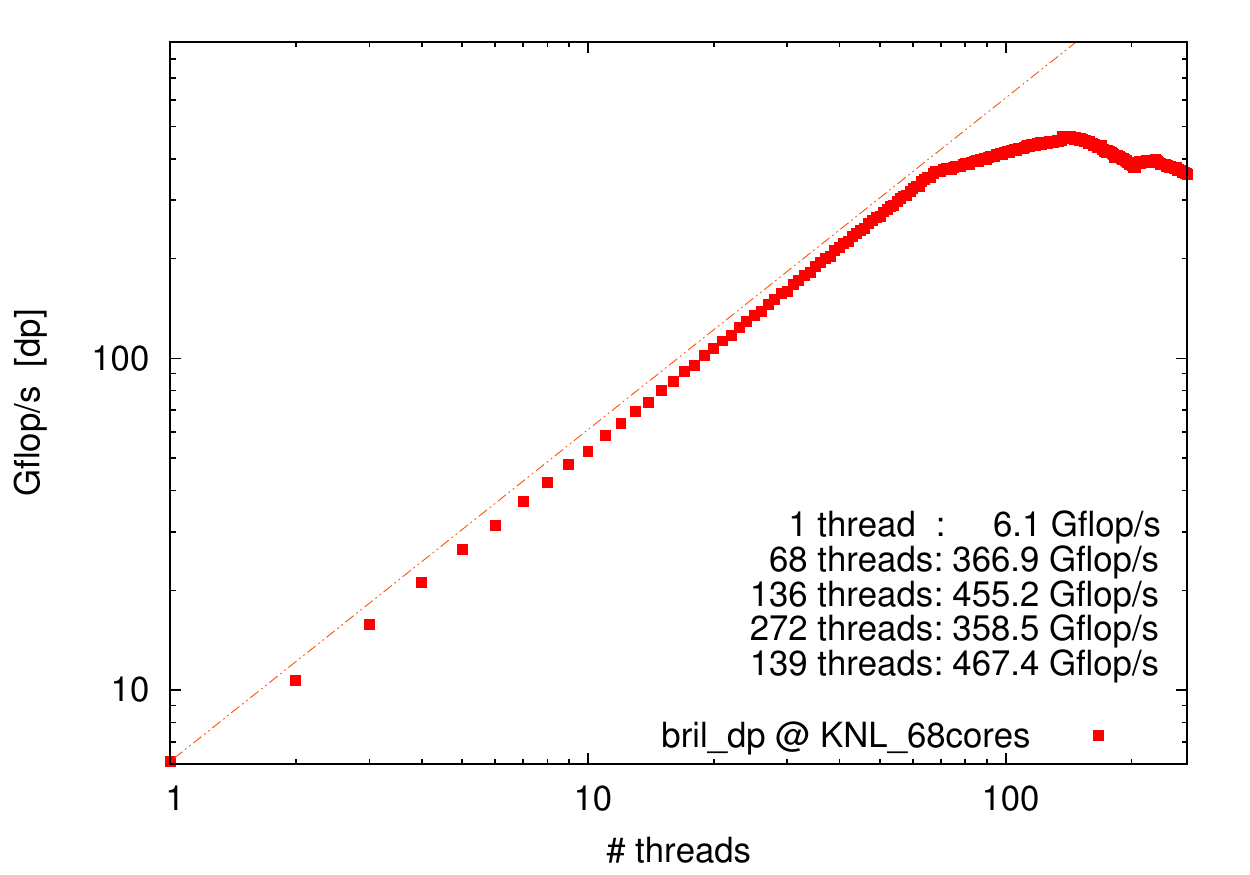}%
\\[2mm]
\includegraphics[width=0.5\textwidth]{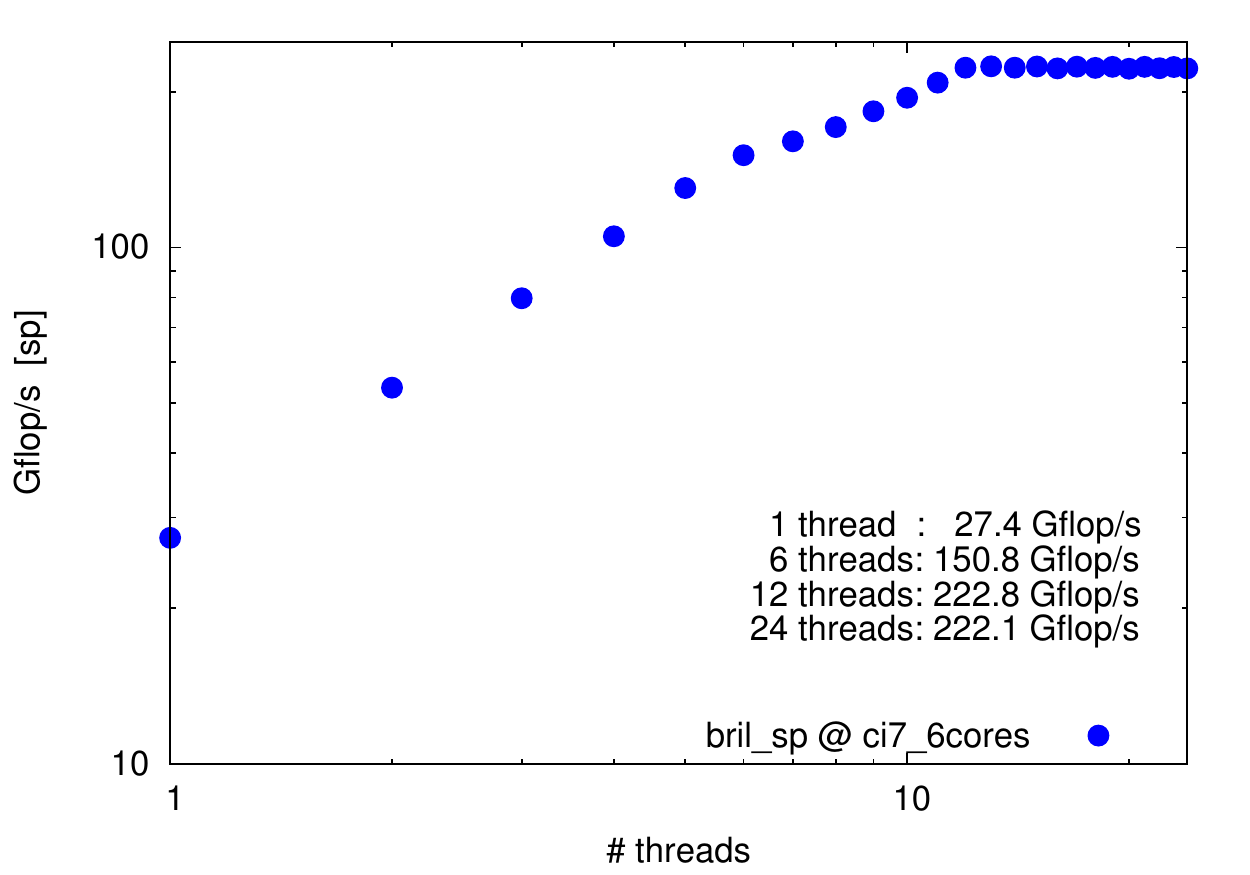}%
\includegraphics[width=0.5\textwidth]{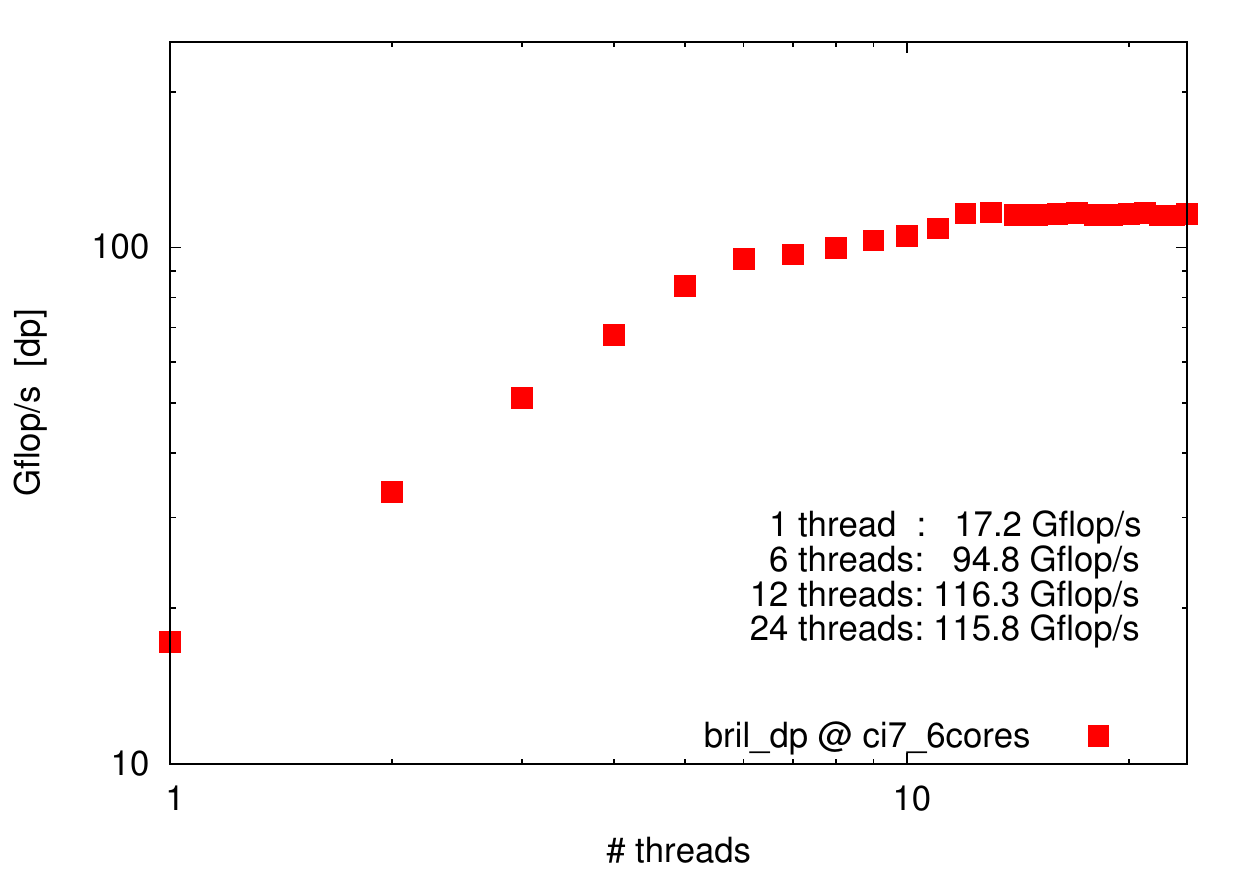}%
\caption{\label{fig:bril}
Log-log scaling plot of the single processor performance in Gflop/s versus the number of OMP threads for the Brillouin Dirac operator at $c_\mr{SW}=0$ in {\tt sp} (left) and {\tt dp} (right).
The top panels feature a KNL processor with 68 cores, the bottom panels a Broadwell chip with 6 cores. The lattice size is $34^3\times68$.}
\end{figure}

The Brillouin operator (\ref{def_bril}) brings new perspectives on PDFs~\cite{Durr:2010ch,Durr:2012dw}
and --~when used in conjunction with the overlap procedure~\cite{Neuberger:1997fp,Neuberger:1998wv}~-- on heavy-quark physics~\cite{Cho:2015ffa,Durr:2017wfi}.
From a HPC viewpoint the Brillouin operator is interesting, since its computational intensity is by a factor $2.5$ higher than for the Wilson operator.
The choice of treating $N_v$ right-hand-sides simultaneously enhances the computational intensity of either operator, while keeping this ratio almost invariant~\cite{Durr:2017wfi}.
In a very distant future, when cycles are totally irrelevant, one might opt for \emph{not} precomputing $W$; this would trigger a massive enhancement of the computational intensity of the operator (\ref{def_bril}).

Thread scaling results are shown in Fig.\,\ref{fig:bril}.
As in previous cases, we see nearly-perfect scaling behavior up to 68 threads on the KNL.
After this, there is another performance increase up to 136 threads.
Beyond that point, performance figures wiggle a bit out to 272 threads.
And again, the (unchanged, just recompiled) code performs reasonably well on the Broadwell architecture, too.
Unlike in the Wilson case, the author is unaware of simple recipes for further improvement.


\section{Summary and outlook}

In summary, thread scaling results for the Susskind, Wilson and Brillouin varieties of the Dirac operator in lattice QCD were presented.
They are based on straightforward implementations in Fortran\,2008, with OpenMP pragmas for shared-memory parallelization and SIMD pipelining.

No cache-access optimization and no hand-assembly tuning have been applied, and still reasonable performance figures can be obtained.
Key to the improvement over last year's version~\cite{Durr:2017clx} is an ansatz where performance critical routines are written for a variety of layouts of the in/out-vectors, of accumulation variables, and possibly loop nestings.
For a given architecture and compiler combination, all of these routines are compiled ``out of the box'', and a few test calls will quickly reveal which option features best on a particular machine.
With this ``best of breed'' ansatz, the winning combination is subsequently used for actual computations.

The plots presented in this contribution illustrate that this strategy proves successful on the single-node level.
What would be important for actual calculations, however, is a working concept (along these lines) on the multi-node level.
The author's hope is that vendors will finally provide efficient PGAS (Coarray Fortran and/or Unified\,Parallel\,C) support for CPUs.
In fact, this is the rationale for not sacrificing any of the space-time indices for the SIMD vectorization.
In the event the GPU world offers this feature more promptly, this will be a clear case for switching to OpenACC.

{\bf Acknowledgements}:
Program development and test runs were performed on the DEEP-ER system at IAS/JSC in J\"ulich.
The author likes to thank Eric Gregory for useful discussion.




\begin{thebibliography}{99}

\bibitem{Boyle:2017wul} 
  P.~A.~Boyle,
  PoS LATTICE {\bf 2016}, 013 (2017)
  doi:10.22323/1.256.0013
  [arXiv:1702.00208 [hep-lat]].
\bibitem{Rago:2017pyb} 
  A.~Rago,
  EPJ Web Conf.\  {\bf 175}, 01021 (2018)
  doi:10.1051/epjconf/201817501021
  [arXiv:1711.01182]. 
\bibitem{Lin:2018}
  M.~Lin,
  ``Machines and Algorithms for Lattice QCD,''
  talk at Lattice 2018 (these proceedings).

\bibitem{DeTar:2016ndn} 
  C.~DeTar, D.~Doerfler, S.~Gottlieb, A.~Jha, D.~Kalamkar, R.~Li and D.~Toussaint,
  PoS LATTICE {\bf 2016}, 270 (2016)
  doi:10.22323/1.256.0270
  [arXiv:1611.00728 [hep-lat]].
\bibitem{DeTar:2018pyj} 
  C.~DeTar, S.~Gottlieb, R.~Li and D.~Toussaint,
  EPJ Web Conf.\  {\bf 175}, 02009 (2018).
  doi:10.1051/epjconf/201817502009
  [arXiv:1712.00143 [hep-lat]].

\bibitem{Durr:2010ch}
  S.~Durr and G.~Koutsou,
  Phys.\ Rev.\ D {\bf 83}, 114512 (2011)
  [arXiv:1012.3615 [hep-lat]].
\bibitem{Durr:2012dw} 
  S.~Durr, G.~Koutsou and T.~Lippert,
  Phys.\ Rev.\ D {\bf 86}, 114514 (2012)
  [arXiv:1208.6270 [hep-lat]].

\bibitem{Neuberger:1997fp}
  H.~Neuberger,
  Phys.\ Lett.\ B {\bf 417}, 141 (1998)
  [hep-lat/9707022].
\bibitem{Neuberger:1998wv} 
  H.~Neuberger,
  Phys.\ Lett.\ B {\bf 427}, 353 (1998)
  [hep-lat/9801031].

\bibitem{Cho:2015ffa}
  Y.~G.~Cho, S.~Hashimoto, A.~Juttner, T.~Kaneko, M.~Marinkovic, J.~I.~Noaki and J.~T.~Tsang,
  JHEP {\bf 1505}, 072 (2015)
  [arXiv:1504.01630 [hep-lat]].
\bibitem{Durr:2017wfi}
  S.~Durr and G.~Koutsou,
  arXiv:1701.00726 [hep-lat].
\bibitem{Durr:2017clx} 
  S.~Durr,
  EPJ Web Conf.\  {\bf 175}, 02001 (2018)
  doi:10.1051/epjconf/201817502001
  [arXiv:1709.01828]. 

\end{thebibliography}
\end{document}